# Reply to “Comment on ‘Dark pulse emission of a fiber laser’”

**Han Zhang, Dingyuan Tang*, Luming Zhao and Xuan Wu**

*School of Electrical and Electronic Engineering, Nanyang Technological University, Singapore 639798*

**Corresponding author: edytang@ntu.edu.sg*

We reply to S. Coen and T. Sylvestre’s comment on our paper [Phys. Rev. A 80, 045803 (2009)] and make some additional remarks on our experimental results.

In their comment on our paper [1], S. Coen and T. Sylvestre pointed out that the dark pulses shown in Fig. 2(a) of our paper could not be 8 ps as it would not be detectable with a 2 GHz bandwidth photo-detector [2]. They could be partially correct. In our subsequent experimental studies on the dark pulse emission of the laser, we further identified that there are in fact two types of dark pulses formed in the laser: one is the Nonlinear Schrödinger Equation (NLSE) type dark solitons [2] and the other one is the optical domain wall (DW) type dark solitons [3, 4]. Depending on the laser operation conditions, which can be changed by tuning the intracavity polarization controller, both types of the dark solitons can coexist in the laser. The DW type dark solitons have lower formation threshold and broader FWHM pulse width than those of the NLSE type dark solitons. However, formation of the DW type dark solitons requires that the laser oscillate simultaneously at two wavelengths. Due to their incoherent nonlinear coupling, the laser emission switches between the two wavelengths, leading to the formation of domain wall dark solitons [4]. When the laser oscillates at single wavelength, only the NLSE type dark solitons could be obtained. The upper trace of Fig. 2(a) in [2] could be the domain wall solitons, while the lower trace of Fig. 2(a) was the case where multiple NLSE type dark solitons coexist in the cavity.

Obviously, different NLSE type dark solitons formed in the laser have different darkness. Most probably, their pulse widths were also different. As the measured optical spectrum is an average of all these solitons, the narrowest dark solitons in the laser would be ~ 8 ps as estimated from the measured optical spectral bandwidth. We were unclear how Coen and Sylvestre calculated the depth-to-background intensity ratio and came to the conclusion that 8 ps dark pulse is undetectable with a 2 GHz detector. Given that their calculation was correct, it would not exclude the existence of 8 ps dark soliton in the laser but confirm the experimental results that NLSE dark solutions with different pulse widths could be simultaneously formed in the fiber laser.